\documentstyle[amssymb,twoside,fleqn,espcrc2]{article}
\makeatletter
\def\eqnarray{\stepcounter{equation}\let\@currentlabel=\theequation
\global\@eqnswtrue
\global\@eqcnt\z@\let\\=\@eqncr
\abovedisplayskip\topsep\ifvmode\advance\abovedisplayskip\partopsep\fi
\belowdisplayskip\abovedisplayskip
\belowdisplayshortskip\abovedisplayskip
\abovedisplayshortskip\abovedisplayskip
$$\m@th\halign
to\linewidth\bgroup\@eqnsel\hskip\@centering$\displaystyle\tabskip\z@
  {##}$&\global\@eqcnt\@ne \hfil${##}$\hfil
  &\global\@eqcnt\tw@ $\displaystyle{##}$\hfil
   \tabskip\@centering&\llap{##}\tabskip\z@\cr}
\def\endeqnarray{\@@eqncr\egroup
      \global\advance\c@equation\m@ne$$\global\@ignoretrue
      }
\def\@maketitle{
  \global\setbox\fm@box=\vbox\bgroup
  \hbox to \hsize {\hss IFUP-TH 45/96}
    \vskip 2mm
    \raggedright                  
    \hyphenpenalty\@M             
    {\Large \@title \par}         
    \vskip\@bls                   
    {\normalsize                  
     \@author \par}               
    \vskip\@bls                   
    \@address                     
  \egroup
  \twocolumn[
    \unvbox\fm@box                
    \vskip\@bls                   
    \unvbox\abstract@box          
    \vskip 2pc]}                  
\makeatother
\title{Critical behavior of the correlation function of
three-dimensional ${\rm O}(N)$ models in the symmetric phase}
\author{Massimo Campostrini, Paolo Rossi, Andrea Pelissetto,
and Ettore Vicari,\\
Dipartimento di Fisica dell'Universit\`a and I.N.F.N., 
I-56126 Pisa, Italy}

\begin{document}

\begin{abstract}
We present new strong-coupling series for ${\rm O}(N)$ spin models in
three dimensions, on the cubic and diamond lattices. We analyze these
series to investigate the two-point Green's function $G(x)$ in the
critical region of the symmetric phase. This analysis shows that the
low-momentum behavior of $G(x)$ is essentially Gaussian for all $N$
from zero to infinity. This result is also supported by a large-$N$
analysis.
\end{abstract}

\maketitle

\section{INTRODUCTION}

Three-dimensional ${\rm O}(N)$-symmetric spin models describe many
important critical phenomena in nature: the case $N=3$ describes
ferromagnetic materials, where the order parameter is the
magnetization; the case $N=2$ describes the helium superfluid
transition, where the order parameter is the quantum amplitude; the
case $N=1$ (Ising model) describes liquid-vapor transitions, where the
order parameter is the density.

The critical behavior of the two-point correlation function $G(x)$ is
related to critical scattering, which is observed in many
experiments, e.g., neutron scattering in ferromagnetic materials,
light and X-rays scattering in liquid-gas systems.

In the following we will focus on the low-momentum behavior of the
Fourier-transformed correlation function $\widetilde{G}(k)$ in the
critical region of the symmetric phase, i.e., for
\[ 
|k|\lesssim 1/\xi, \qquad 0<T/T_c - 1\ll 1.
\] 

\section{LATTICE MODELS}

Let us consider an ${\rm O}(N)$-symmetric lattice spin models
described by the nearest-neighbor action
\begin{equation}
S = -N\beta \sum_{\rm links}  \vec{s}_{x_l}\cdot \vec{s}_{x_r} \,,
\end{equation}
where $\beta=1/T, $ $\vec{s}$ is an $N$-component real vector, and
$x_l$, $x_r$ are the endpoints of the link.  The two-point correlation
function is defined by
\begin{equation}
G(x)=\langle \vec{s}_x \cdot \vec{s}_0 \rangle .
\end{equation}

In order to simplify the study the critical behavior of $G(x)$, we
introduce the dimensionless RG-invariant function
\begin{equation}
L(k;\beta)\equiv{\widetilde{G}(0;\beta)\over\widetilde{G}(k;\beta)}\,.
\end{equation}
In the critical region of the symmetric phase, $L(k,\beta)$ is a
function only of the ratio $y\equiv k^2/M_G^2$, where 
$M_G\equiv 1/\xi_G$; the second-moment correlation length $\xi_G$ is 
defined by
\begin{equation}
\xi_G^2\equiv {1\over 6}
{ \sum_x x^2 G(x)\over \sum_x G(x) }.
\end{equation}
$M_G$ is the mass-scale which can be directly observed in scattering
experiments.
$L(y)$ can be expanded in powers of $y$ around $y=0$:
\begin{equation}
L(y) = 1 + y + l(y), \qquad
l(y) = \sum_{i=2}^\infty c_i y^i.
\end{equation}
$l(y)$ parameterizes the difference from a generalized Gaussian
propagator.  The coefficients $c_i$ can be expressed as the critical
limit of appropriate dimensionless RG-invariant ratios of the
spherical moments 
\begin{equation}
m_{2j} = \sum_x x^{2j} G(x).
\end{equation}

Another interesting quantity related to the low-momentum behavior of
$G$ is the ratio $s=M^2/M_G^2$, where $M$ is the mass-gap of the
theory.  Its critical value is $s^*=-y_0$, where $y_0$ is the zero of
$L(y)$ closest to the origin.

In the large-$N$ limit, $l(y)$ is depressed by a factor of $1/N$.  The
coefficients $c_i$ can be obtained from a $1/N$ expansion in the
continuum \cite{Aharony}:
\begin{eqnarray}
&& c_{2}\simeq-{0.0044486 \over N}\,, \quad
c_{3}\simeq{0.0001344 \over N}\,, \nonumber \\
&& c_{4}\simeq-{0.00000658 \over N}\,, \quad
c_{5}\simeq{0.00000040 \over N}\,... 
\end{eqnarray}
We are presently computing the order $1/N^2$ of the expansion.
We expect that the pattern established by the $1/N$ expansion
\begin{equation}
c_i\ll c_2\ll 1, \quad i\geq 3
\end{equation}
will be followed by all models with sufficiently large $N$.
This implies $s^* -1 \simeq c_2$: indeed, in the large-$N$ limit,
\begin{equation}
s^*-1\simeq -{0.0045900 \over N}\,.
\end{equation}

The coefficients $c_i$ can also be computed from an
$\varepsilon$-expansion of the corresponding $\phi^4$ theory around
$d=4$ \cite{Fisher-Aharony}:
\begin{equation}
c_i \simeq \varepsilon^2 {N+2\over (N+8)^2}\, e_i,
\end{equation}
where $\varepsilon=4-d$ and
\begin{equation}
e_2 \simeq -0.007520 ,\qquad e_3\simeq 0.0001919.
\end{equation}

\section{STRONG-COUPLING EXPANSION}

We computed the strong-coupling expansion of $G(x)$ up to 15th order
on the cubic lattice, and up to 21st order on the diamond lattice.
Our technique for the strong-coupling expansion of ${\rm O}(N)$ spin
models was presented in Ref.\ \cite{Melbourne}.

We took special care in the choice of estimators for the ``physical''
quantities $c_i$ and $s^*$. This step is very important from a
practical point of view: better estimators can greatly improve the
stability of the extrapolation to the critical point.  Our search for
optimal estimators was guided by the requirement of a regular
strong-coupling expansion (e.g., no $\ln\beta$ terms) and by the
knowledge of the large-$N$ limit (we chose estimators which are
``perfect'' for $N=\infty$).

The strong-coupling series of the estimators were analyzed by Pad\'e
approximants, Dlog-Pad\'e approximants and first-order integral
approximants (see Ref.\ \cite{Guttmann} for a review of the
resummation techniques; see also Ref.\ \cite{SCN3}).
For diamond lattice models with $N\ne0$, $\beta_c$ was not known, and
we estimated it from the strong coupling series of the magnetic
susceptibility.

\begin{table*}[tb]
\setlength{\tabcolsep}{1.5pc}
\caption{Comparison of strong-coupling expansion on cubic and diamond
lattices with $1/N$ and $\varepsilon$-expansion}
\label{main-table}
\renewcommand\arraystretch{1.2}
\begin{tabular}{ccr@{}lr@{}lr@{}l}
\hline
\multicolumn{1}{c}{$N$}&
\multicolumn{1}{c}{lattice}&
\multicolumn{2}{c}{$10^{4} c_2$}&
\multicolumn{2}{c}{$10^{5} c_3$}&
\multicolumn{2}{c}{$10^{4} (s^*-1)$}\\
\hline 
$0$ & cubic   & \multicolumn{2}{c}{$|10^{4} c_2|\lesssim 2$}
    & $1.$&$2(1)$ & $1.$&$2(3)$ \\
    & diamond & \multicolumn{2}{c}{$|10^{4} c_2|\lesssim 1$}
    & $1.$&$0(1)$ & $1.$&$0(5)$ \\
    & $\varepsilon$-expansion & $-2.$&$35$ & $0.$&$60$ & & \\\hline
$1$ & cubic   & $-2.$&$9(2)$ & $1.$&$1(1)$ & $-2.$&$3(5)$ \\
    & diamond & $-3.$&$1(2)$ & $1.$&$0(2)$ & $-2.$&$2(3)$ \\
    & $\varepsilon$-expansion & $-2.$&$78$ & $0.$&$71$ & & \\\hline
$2$ & cubic   & $-3.$&$8(3)$ & $1.$&$1(1)$ & $-3.$&$5(5)$ \\
    & diamond & $-4.$&$2(3)$ & $1.$&$1(3)$ & $-3.$&$5(2)$ \\
    & $\varepsilon$-expansion & $-3.$&$01$ & $0.$&$77$ & & \\\hline
$3$ & cubic   & $-4.$&$0(2)$ & $1.$&$1(2)$ & $-4.$&$0(4)$ \\
    & diamond & $-4.$&$2(3)$ & $1.$&$1(3)$ & $-3.$&$5(2)$ \\
    & $\varepsilon$-expansion & $-3.$&$11$ & $0.$&$79$ & & \\\hline
$4$ & cubic   & $-4.$&$1(2)$ & $1.$&$2(1)$ & $-4.$&$0(4)$ \\
    & diamond & $-4.$&$7(2)$ & $1.$&$0(2)$ & $-4.$&$0(2)$ \\
    & $\varepsilon$-expansion & $-3.$&$13$ & $0.$&$80$ & & \\
    & $1/N$   & $-11.$&$12$ & $3.$&$36  $ & $-11.$&$48$ \\\hline
$8$ & cubic   & $-3.$&$5(2)$ & $1.$&$0(2)$ & $-3.$&$7(3)$ \\
    & diamond & $-4.$&$0(1)$ & $0.$&$7(5)$ & $-4.$&$0(4)$ \\
    & $\varepsilon$-expansion & $-2.$&$94$ & $0.$&$75$ & & \\
    & $1/N$   & $-5.$&$56 $ & $1.$&$18 $ & $-5.$&$74 $ \\\hline
$16$& cubic   & $-2.$&$4(2)$ & $0.$&$70(5)$& $-2.$&$7(2)$ \\
    & diamond & $-2.$&$65(5)$ & $0.$&$5(5)$ & $-2.$&$9(2)$ \\
    & $\varepsilon$-expansion & $-2.$&$35$ & $0.$&$60 $ & & \\
    & $1/N$   & $-2.$&$78 $   & $0.$&$84  $ & $-2.$&$87  $ \\\hline
\end{tabular}
\end{table*}

Our strong-coupling results on cubic and diamond lattices are compared
with the results of the $1/N$ expansion and of the
$\varepsilon$-expansion in Table \ref{main-table}.  One may notice
that universality between cubic and diamond lattice is always
confirmed; furthermore, the agreement with the $\varepsilon$-expansion
and with the $1/N$ expansion is satisfactory.

The predicted pattern $c_3\ll c_2\ll 1$ is verified for all $N$.  We can
conclude that the two-point Green's function is essentially Gaussian
for all momenta with $|k^2|\lesssim M_G^2$, and that the small
corrections are dominated by the $(k^2)^2$ term.

\section{APPROACH TO CRITICALITY}

We investigated the approach to criticality, with special
attention devoted to anisotropy (violation of rotational invariance).
Let us introduce the anisotropy estimators
\begin{eqnarray}
&& l_4 = \sum_{x,y,z} [f_4(x,y)+f_4(y,z)+f_4(z,x)] \, G(x,y,z),
   \nonumber \\
&& \quad f_4(x,y) = (x^2+y^2)^2 - 8x^2y^2;  \\
&& l_{6,1} = \sum_{x,y,z} [f_6(x,y)+f_6(y,z)+f_6(z,x)] \nonumber \\ 
&& \qquad \times G(x,y,z), \nonumber \\
&& \quad f_6(x,y) = (x^2+y^2)^3 - 8(x^4y^2+x^2y^4);   \\
&& l_{6,2} = \sum_{x,y,z} [x^6+y^6+z^6 - 45x^2y^2z^2]\, G(x,y,z).
   \nonumber \\
\end{eqnarray}
In the critical limit, $l_{2j}$ are depressed 
with respect to the spherical moments $m_{2j}$.
In the large-$N$ limit one can show that
\begin{equation}
A_{2j,i}\equiv { l_{2j,i}\over m_{2j} }\sim \xi_G^{-2}.
\label{approach}
\end{equation}
We analyzed the strong-coupling series of
\begin{equation}
B_{2j,i}\equiv {l_{2j,i}\over m_{2j-2}}\,;
\end{equation}
for all values of $N$, we found that $B_{2j,i}$ have a finite (but
non-universal) $T\to T_c$ limit.  This supports the validity of Eq.\
(\ref{approach}) for all $N$.

Ratios of $A_{2j,i}$ are universal quantities; we found that at
criticality $A_{6,1}/A_4\simeq 0.95$ and $A_{6,2}/A_{6,1}\simeq 0.75$
(within one per mill) for all $N$.

\end{document}